\documentclass[11pt,a4paper]{article}
\usepackage{braket} 
\usepackage{amsmath}                      
\usepackage{amsfonts}                    
\usepackage{amssymb}                     
\usepackage{amsthm}                       
\usepackage{indentfirst}                  
\usepackage{geometry}                    
\usepackage{eufrak}                       
 \usepackage{hypbmsec} 
 \usepackage{jheppub}                    
 \usepackage{graphicx}                    
 \usepackage{epstopdf}                    
 \usepackage{caption}                     
 \usepackage{bbm}

\graphicspath{{figures/}}

\numberwithin{equation}{section}
\numberwithin{figure}{section}

\begin{document}

\title{\Huge Celestial CFT from CHY Formalism: Center Charge and Finite Size Effect\\
}

\author{Ming Yu}
\date{\today}
\affiliation{CAS Key Laboratory of Theoretical Physics, Institute of Theoretical Physics, Chinese Academy of Sciences,  Beijing 100190, China}
\emailAdd{yum@itp.ac.cn}
\abstract{Scattering amplitudes in gauge theories can be calculated either by bulk theories in 4d Minkowski space-time($Mink_4$), or perceived as the correlation functions in celestial CFT(CCFT) living in the celestial sphere at null infinity, where  an infinite-dimensional asymptotic symmetry, BMS group,  resides.
	Another well developed method is the CHY formalism, which  formulates the scattering amplitude in terms of the correlation functions on a 2d world sheet, on which an ambitwistor string theory  is defined. The relationship between CHY theory and  CCFT is encoded in scattering equations, which are  algebraic equations lacking of analytical solutions in general. So we start from the CHY formalism, take the collinear limit, then find a nice operator formalism for the CCFT. In particular, the center charge $c$ is calculated to be 36  for the CCFT related to the  4d Yang-Mills theory. It then follows that the 4d cosmological constant naturally  arises as the finite size effect in 2d CCFT, which is calculated by the method of  $T\bar{T}$ perturbed CCFT.
}


\maketitle
\flushbottom

\newpage\clearpage{\pagestyle{empty}\cleardoublepage}


\newcommand{\dif}{\mathrm{d}} 
\newcommand{\mi}{\mathrm{i}}  
\newcommand{\me}{\mathrm{e}}  

\section{Introduction}
Scattering amplitudes in gauge theories can be calculated either by bulk theories in 4d Minkowski space-time($Mink_4$), or considered as the correlation functions in celestial CFT living in the celestial sphere at null infinity, where  an infinite-dimensional asymptotic symmetry resides.
This asymptotic  symmetry group is originally  called BMS symmetry group \cite{bms}. Later, these symmetry algebra is enlarged to include analytically  singular transformations in  celestial sphere \cite{bt, banks}, hence called the extended BMS symmetry. The extended BMS symmetry is, quantum mechanically, the
Virasoro and Kac-Moody type of symmetry in  celestial CFT(CCFT) \cite{stronminger1308.0589, Strominger1406.3312, Strominger1503.02663, Strominger1609.00282}.
Another well developed method is the CHY formalism \cite{cachazo2014scattering, cachazo2013scattering}, which  formulates the scattering amplitude in terms of the correlation functions on a 2d world sheet on which an ambitwistor string theory \cite{mason2013ambitwistor} is defined. The relation between CHY theory and the CCFT is encoded in scattering equations, which are  algebraic equations lacking of analytical solutions. This relation is further complicated by the fact that while there is an explicit construction for the string vertex operator in the formal case, there is none for the later at the moment. Thus a compelling questions is: can we construct the  energy momentum tensor in CCFT in terms of the worldsheet fields in ambitwistor string theory, in analogy  to the $AdS_3/CFT_2$  formalism \cite{seiberg9806194,yum9812216}?  As the subject is developed in this paper, we shall see that in the collinear limit, we indeed find a nice operator formalism for the CCFT. In particular, the center charge, and the cosmological constant, which arises as the finite size effect in CCFT,  are calculated.

BMS symmetry is an infinite-dimensional symmetry for the scattering amplitude of massless particles, exhibiting an analytical structure emerging from the asymptotically flat space in the past and the future, see \cite{Frolov} for a  review. According to the authors of ref.\cite{stronminger1308.0589, Strominger1406.3312}, the BMS symmetry is realized
as a 2d conformal symmetry in the celestial space,
with the decomposition of the three dimensional null momentum into
two dimensional conformal coordinate space and one dimensional
conformal weight space. Indeed, in spinor space, the so(3,1) Lorentz transformation is
explicitly decomposed as $su(2,c)$ global conformal symmetry in 2d
space time, $k^\mu=\omega q^\mu$, 
$q^\mu=\frac{1}{2}\begin{pmatrix}u& 1\end{pmatrix}\sigma^\mu
\begin{pmatrix}\bar{u}\\ 1\end{pmatrix}=\frac{1}{2}\mathrm{Tr}\big(\sigma^\mu \begin{pmatrix}\bar{u}\\ 1\end{pmatrix}\begin{pmatrix}u& 1\end{pmatrix}\big)=\frac{1}{2}\begin{pmatrix}u\bar{u}+1,& u+\bar{u},& -i(u-\bar{u}),&u\bar{u}-1\end{pmatrix}$. Hence, we can write $\begin{pmatrix}\bar{u}\\ 1\end{pmatrix}\begin{pmatrix}u & 1\end{pmatrix}=q^\mu \bar{\sigma}_\mu$. Here,  $\sigma^i$'s ($\bar{\sigma}_i$'s) are the Pauli matrices, $\sigma^0$ ($\bar{\sigma}_0$) =$\mathbbm{1}$.
One can also define the following null vectors for later use,
$\epsilon^+=\partial_u q,\ \epsilon^-=\partial_{\bar{u}} q,\  \mu=\partial_u \partial_{\bar{u}}q$.	
The three
independent variables, $\omega, u, \bar{u}$ can be viewed as the
celestial coordinates on the asymptotically past and future infinity, on
which infinitely many asymptotic symmetries, called BMS symmetry,
are expected to exist.
Exactly because of the existence of the continuous spectrum of
the conformal weights,  it is difficult to extract  useful information from
some particular examples of the celestial conformal field
theory. In refs.\cite{Strominger1701.00049, Strominger1706.03917, fan2103.04420, Strominger2104.13432}, efforts have been made in extracting conformal
block formalism starting from  the scattering amplitudes of few particles.
Yet, it remains unclear how to generalize them to the operator
formalism realizing the  infinitely dimensional conformal algebra at
the level of scattering amplitudes of massless particles.

One may ask if the celestial CFT is the holographic dual to the 4D 
quantum field theory in Minkowski space time ($Mink_4$). One way to see that relation is to foliate 
$Mink_4$ into warped $AdS_3$ slices, then the celestial  space is just the slices of the $AdS_3$
boundaries \cite{Solodukhin0303006, Solodukhin0405252, Cheung, Strominger1905.09809, Strominger2204.10249, Strominger2312.07820}. This point of view will be proved useful when considering the finite size effect of the celestial CFT.
To foliate $Mink_4$ into slices of the warped spaces, we may consider a Euclidean subspace of $Mink_4$ satisfying
\begin{align*}
x^2=-e^{2\tau}
\end{align*} 
This subspace is called Milne space($Miln_4$). $x^\mu$ can be parametrized as  
\begin{align*}
x^\mu=e^\tau(\frac{q^\mu}{\rho}+\rho v^\mu).
\end{align*}

Another apparent different approach is the  CHY formalism, which resembles the string theory description of the target space quantum filed theory.  In CHY formalism, the string vertex operators $\mathcal{V}_k(z_i)$'s are
  inserted on a 2-dimensional surface.  The coordinate $z_i$'s is to be integrated around the coordinate $\hat{z_i}$, which are related to the null vector $k_i$'s through the so called scattering equations,
\begin{equation}\label{se}
	\sum_{j\neq i} \frac{k_i\cdot k_j}{\hat{z}_i-\hat{z}_j}=0, \ \ i=1,2,...,n.
\end{equation}
If we call the $\hat{z}_i$'s  the worldsheet coordinates, on which the CHY formalism is defined, then the implied string theory is a
particular degenerate string theory, the so called ambitwistor string
theory, which is obtained from the ordinary string theory by taking
the $\alpha'$ $\rightarrow \infty$ limit \cite{lizzi1986quantization, gamboa1990null, lindstrom1991zero, lindstrom19303173, siegel2015amplitudes} while choosing a reversed
vacuum state for the anti-holomorphic oscillators \cite{casali2016null}. Effectively, in
this limit, only the zero mass modes survives, so the formalism is
identical to the $\alpha'$ $\rightarrow 0$ limit of the ordinary
string theory.

But what makes the ambitwistor string theory different from the
ordinary string theory is that while the Koba-Nielsen variables,
$z_i$'s,  in the later case are to be integrated along cycles of
macro-scale sizes, the analogue integrations in the former case are
localized along tiny circles of infinitesimal sizes centered in
points on the holomorphic world sheet. In fact the worldsheet
coordinates on which the ambitwistor string vertex operators resides are
related to the celestial coordinates represented by the $k_i$'s by
the so called scattering equation in CHY formalism, eq.(\ref{se})

Because of the sl2 invariance, the solution to equation (1.1) is
$(n-3)!$ fold. So for fixed $k_i$'s, we should sum over $(n-3)!$
different coordinate sets of  vertex operator insertions. The summation over different solutions 
to the scattering equation is strongly reminiscent of the conformal block solutions in 2d CFT's, where summation 
is made over different factorized holomorphic and anti-holomorphic conformal blocks. Since the scattering amplitude can be described 
equivalently in terms of either CCFT or CHY formalism, it is desirable to check the equivalence explicitly.
In the case of $AdS_3$, the Virasoro algebra living on the boundary of the $AdS_3$ space generates the space 
time conformal symmetry, albeit they can be constructed using fields living on the worldsheet.
Here, we way ask a similar question, can we construct the energy momentum tensor in CCFT in
terms of the fields in   ambitwistor string theory?  Naively,  the answer is no, because the
solutions to the scattering equation are algebraically complicated, it is not so straightforward 
to interpret the 2d CFT in celestial space as an ambitwistor string
theory through coordinate transformations. The situation simplifies, however, when
we start with the degenerating cases, i.e. the
cases of the boundary solutions which arise when two vertex
operators are approaching each other. Hopefully in these limits, the
scattering equations are easier to solve, perhaps, recursively. In
that case we might be able to construct the energy momentum tensor in celestial space
quantum mechanically, confirming the scattering amplitude be the
correlation function  in a 2d CFT on the celestial sphere. So the key point in
the present paper is constructing an energy momentum tensor relevant
in celestial CFT, starting with the CHY formalism.

The importance of the construction of the 2d celestial CFT associated with the 4d QFT can be understood in the following way.
First, from the point of view of the 2d CFT, 4d gravity and Yang-Mills theory are treated in an unified way. Second, conformal anomaly
can be calculated explicitly once the 2d energy momentum tensor  is constructed explicitly. Third the conformal anomaly is related to the vacuum energy
of the 4d system and can be related to the 4d cosmological constant through finite size effect. This is a relatively simple way interpreting the dark energy, which is known to exist
in our present universe and is hard to calculate from the  point of view of zero-point energy in 4d QFT.

The main purpose of the present paper is to construct an explicit
quantum operator formalism, which naturally forms a representation
of the asymptotic symmetry algebra and in which, 2d conformal field
theory plays  an essential role. 
 
 In sec.[\ref{ope}], we lay out the basic ingredients in forming CHY formalism for the Yang-Mills theories. In operator formalism, it is the same as the ambitwistor string theory living  in a holomorphic coordinate system on a 2d worldsheet. In doing so, the correlation function
  is just  constructed by Wick contraction of all the oscillating modes of the free fields. The free field realization of the vertex operator is associated with a contour integral around a tiny cycle on the worldsheet. In sec.[\ref{cl}] we then realize that in the collinear limit the contour integration can be carried out and we are left with a vertex operator composed of quasi-local fields. By quasi-local we mean some local fields are sitting in the denominator.    
Since we have simplified the vertex operator inserted at $z_1$ in the collinear limit, it is straightforward to construct the energy momentum tensor in this coordinate patch. This is done in sec.[\ref{emt}]. The center charge is calculated to be $c=11d+d/2+c_g-41$. Here,
$d$ is the dimension of the bulk flat spacetime, $c_g$ the center charge for the Sugawara construction of the stress tensor for the Kac-Moody current algebra. With the center charge in CCFT calculated, we proceed to calculate the finite size effect in sec.[\ref{fse}]. We thus find that the specific value is comparable with that of the  4d cosmological constant. Finally, in sec.[\ref{c}],
we summarize what we have done in the present paper and speculate what could be done in our future work.

\section{OPE}\label{ope}
The short distance operator product expansion (OPE) plays a central role in constructing a 2d CFT. In \cite{Strominger1910.07424, Strominger2105.00331}, Celestial OPE are calculated for gluons and gravitons. 
See also \cite{Sharma2111.02279} for mapping worldsheet OPE to celestial OPE.
Here we take different approaches.  
Since we want to calculate the center charge for CCFT, an explicit operator formalism is needed. So we go to the ambitwistor string formalism. In the collinear limit, we hope that some information on CCFT can be 
obtained via OPE in the explicit operator formalism. 
In what follows we are going to consider  two different OPEs
in different contexts: worldsheet OPE versus spacetime OPE.

In CHY formalism, one can realize the tree level color ordered scattering amplitude in gauge theory  as the correlation function of the n-string
vertex operators inserted on a 2d world sheet.
\begin{align}
S(\{k_i\})&=\sum_{\{Z_I\}}\left\langle V(z_n)U(z_{n-1})U(z_{n-2})\prod_{\hat{z}_i\in Z_I}\oint_{\hat{z}_i}dz_i \mathcal{V}_{k_i}(z_i)\right\rangle
\end{align}
The summation is over different sets of solutions to the scattering equations. Here,  $z_i$'s, i=1,2,...n-3, are sets of points on the 2d worldsheet and are related to
the null momentum of the gauge bosons, $\{k_i\}$'s,  by the  the
scattering equations, eq.(\ref{se}). The integrand of the string vertex operator is
constructed as follows,
\begin{align}
\mathcal{V}_k(z)&=\frac{e^{k\cdot X(z)}}{k\cdot P(z)}\left(\epsilon\cdot P(z)+k\cdot\psi(z)\epsilon\cdot\psi(z)\right)J(z)
\end{align}
Or in terms of 2d super-coordinates,
\begin{align}
	\oint dz \mathcal{V}_k(z)&=\int e^{k\cdot Y(z,\bar{z},\theta)}\epsilon\cdot D Y(z,\bar{z},\theta)J(z)d^2z d\theta\\
	Y(z,\bar{z},\theta)&=X(z)+\theta \psi(z)+\bar{z}P(z)\\
	D&=\partial_\theta+\theta \partial_{\bar{z}}
\end{align}
$X(z), P(z)$ are 4d vector  and 2d holomorphic  bosonic
fields of conformal dimension 0 and 1 respectively, $\psi(z)$ is a 4d vector  and 2d holomorphic conformal dimension
$\frac{1}{2}$ Majorana fermionic fields, $J(z)$ the Kac-Moody currents, $\epsilon$ the polarization 4-vector of the gauge boson. 
$Y(z,\bar{z},\theta)$ is a compact form of the 2d superconformal
field. 
\begin{align}
P^\mu(w)X^\nu(z)&=\frac{\eta^{\mu\nu}}{w-z}\\
\psi^\mu(w)\psi^\nu(z)&=\frac{\eta^{\mu\nu}}{w-z}
\end{align}
with space time signature $(-1,1,1,1)$.

The constraining  superconformal algebra in the present setting reads as
following,
\begin{align}
	G(z)&=P(z)\cdot\psi(z) \\
	B(z)&=P(z)\cdot P(z)\\
	T(z)&=\partial_z X(z)\cdot P(z)+\frac{1}{2} \partial_z\psi(z) \cdot \psi(z)
\end{align}
$G(z)$ and $B(z)$ are the primary fields of conformal weight $3/2$
and 2, respectively, with respect to the energy momentum tensor
$T(z)$.
\begin{align}
	T(w)G(z)&=\frac{\partial_z G(z)}{w-z}+\frac{3}{2}\frac{G(z)}{(w-z)^2}\\
	T(w)B(z)&=\frac{\partial_z B(z)}{w-z}+2\frac{B(z)}{(w-z)^2}\\
	G(w)G(z)&=\frac{B(z)}{w-z}
\end{align}
To fix the global
superconformal symmetry, we can fix three bosonic coordinates,
say, $z_n,\  z_{n-1},\  z_{n-2},$ to fixed values and two fermionic coordinates, say, $\theta_{n-1},\ \theta_{n-2}$ to zero.

Because of the global superconformal gauge invariance of the correlation function, 
we also need two U's and one V which are the non-integrated vertex operators. Ghost fields are inserted 
to absorb the background ghost charges,
\begin{align}
U(z)&=c(z)\tilde{c}(z)\delta(\gamma(z))e^{k\cdot X(z)}\epsilon_i\cdot \psi(z)J(z)	\\
V(z)&=c(z)\tilde{c}(z)e^{k\cdot X(z)}(\epsilon\cdot P(z)+k\cdot\psi(z)\epsilon_i\cdot \psi(z))J(z)
\end{align}	
Here, $c$ and $\tilde{c}$ are the fermionic $(-1)$ form ghost fields, $\gamma$ is bosonic $(-\frac{1}{2})$ form ghost field.
In bosonized form, $\delta(\gamma(z))=e^{-\phi(z)}, \ \ \ <\phi(z) \phi(z^\prime)>=-\log(z-z^\prime), \ \ \ :\beta(z)\gamma(z):=\partial_z\phi(z)$,
Hence we have 
\begin{align}
S(\{k_i\})&=V(z_n)U(z_{n-1})U(z_{n-2})\sum_{\{Z_I\}}\prod_{\hat{z}_i\in Z_I}	\oint_{\hat{z_i}}\frac{e^{k_i\cdot X(z_i)}}{k_i\cdot P(z_i)}(\epsilon_i\cdot P(z_i)+k_i\cdot\psi(z_i)\epsilon_i\cdot\psi(z_i))J(z_i)dz_i\\
&=V(z_n)U(z_{n-1})U(z_{n-2})\sum_{\{Z_I\}}\frac{\prod_{\hat{z}_i\in Z_I}	e^{k_i\cdot X(\hat{z_i})}(\epsilon_i\cdot P(\hat{z_i})+k_i\cdot\psi(\hat{z_i})\epsilon_i\cdot\psi(\hat{z_i}))}{\det\{\partial_{\hat{z}_j} f_i(\{\hat{z_l}\})\}}J(\hat{z_i})\\
f_i(\{z_l\})&=\sum_{j\neq i} \frac{k_i\cdot k_j}{z_i-z_j}
\end{align}

Notice that the integration of the $X$ zero mode would produce the delta function $\delta^{(4)}(\sum_{i=1}^n k_i^\mu)$, therefore we are considering the unstripped amplitude with the integration over $X$ zero modes implicitly implied.

\section{Collinear Limit}\label{cl}

Now let's consider the case when the two vertex operators are approaching
each other, $z_i \to z_j $. The interesting solutions to the scattering equation, eq.(\ref{se}), in this case, are the
ones with collinear momenta, $k_i\cdot k_j \to 0$, which is the special case when the
S-matrix factorizes into sub-processes and   leads to the BCFW recursions \cite{bcfw}. The analytical method used in deriving the
BCFW relation in fact suggests that there is a bigger symmetry in
the underlying theory. One way to visualize such a symmetry is to
consider the on shell condition for the propagator as the operator
product expansion (OPE) in 2d conformal field theory (CFT). A tree
amplitude consists of a summation of relevant diagrams which are
dominated by a subset of diagrams when one of the internal line
going on-shell, i.e. $p^2\to 0$
\begin{equation}
	A_n\to A_{n-m+1}\frac{1}{p^2}A_{m+1}
\end{equation}
Letting the resonance of the internal line analytically depends on a
one complex parameter $z$, then one can show under some conditions,
$A_n$ can be solved recursively by analytical methods, provided that we work
out the following OPE.
\begin{align}
	\mathcal{O}_{k_1,\epsilon_1}(z_1)\mathcal{O}_{k_2,\epsilon_2}(z_2)&\sim \sum_{s=\pm}C_{12s}\mathcal{O}_{ k_1+k_2,\epsilon^s}(z_2)
\end{align}
To work out the OPE $\mathcal{O}_{k_1}(z_1)\mathcal{O}_{k_2}(z_2)$, let us first consider the Jacobian for the delta function part. In the following, summation over $l,i,j \neq 1,2$ is assumed and 
only the  singular terms under $z_1\rightarrow z_2$ are retained,
\begin{align}\label{ja}
\nonumber & \det[A_n] \\ \nonumber
=&a_{11}\cdot A^{(11)}-a_{12}\cdot A^{(12)}-(-)^l a_{1l}\cdot A^{(1l)} \\ \nonumber
=&a_{11}\cdot (a_{22}\cdot A^{(12,12)}+(-)^l a_{2l}\cdot A^{(12,1l)})-a_{12}\cdot( a_{21}\cdot A^{(12,12)}+(-)^l a_{2l}A^{(12,2l)})\\ \nonumber
&-(-)^l a_{1l}\cdot (a_{21}A^{(12,1l)}-a_{22}A^{(12,2l)}-(-)^k a_{2k}A^{(12,kl)})\\ \nonumber
=&(a_{11}a_{22}-a_{12}a_{21})\cdot A^{(12,12)}+(-)^l ( a_{11}a_{2l}-a_{1l}a_{21})\cdot A^{(12,1l)})-(-)^l (a_{12}a_{2l}-a_{1l}a_{22}) A^{(12,2l)})\\ \nonumber
=&\left((\frac{s_{12}}{(z_1-z_2)^2}+\frac{s_{1i}}{(z_1-z_i)^2})(\frac{s_{21}}{(z_2-z_1)^2}+\frac{s_{2j}}{(z_2-z_j)^2})-\frac{s_{12}}{(z_1-z_2)^2}\frac{s_{21}}{(z_2-z_1)^2}\right)\frac{A^{(12,12)}}{4}\\ \nonumber
&+(-)^l\left(-(\frac{s_{12}}{(z_1-z_2)^2}+\frac{s_{1i}}{(z_1-z_i)^2})\frac{s_{2l}}{(z_2-z_l)^2}-\frac{s_{1l}}{(z_1-z_l)^2}\frac{s_{21}}{(z_2-z_1)^2}\right) \frac{A^{(12,1l)})}{4}\\ \nonumber
&-(-)^l (\frac{s_{12}}{(z_1-z_2)^2}\frac{s_{2l}}{(z_2-z_l)^2}+\frac{s_{1l}}{(z_1-z_l)^2}(\frac{s_{21}}{(z_2-z_1)^2}+\frac{s_{2j}}{(z_2-z_j)^2})\cdot \frac{A^{(12,2l)})}{4}\\ \nonumber
=&\frac{s_{12}}{4(z_1-z_2)^2}\left(\frac{s_{1i}+s_{2i}}{(z_2-z_i)^2}\cdot A^{(12,12)}-(-)^l \frac{s_{1l}+s_{2l}}{(z_2-z_l)^2}\cdot A^{(12,1l)})-(-)^l \frac{s_{1l}+s_{2l}}{(z_2-z_l)^2}\cdot A^{(12,2l)}\right)\\ \nonumber
=&-\frac{s_{12}}{2(z_1-z_2)^2}\left(-\frac{k\cdot k_i}{(z_2-z_i)^2}\cdot A_{n-1}^{(1,1)}+(-)^l \frac{k\cdot k_l}{(z_2-z_l)^2}\cdot (A^{(12,1l)}+A^{(12,2l)})\right)\\ 
=&-\frac{k_1\cdot k_2}{(z_1-z_2)^2}\det[A_{n-1}]
\end{align}
Here, $k=k_1+k_2$
\begin{align*}
a_{n-1}^{11}=&-\sum_i\frac{k\cdot k_i}{(z_2-z_i)^2}\\
a_{n-1}^{1l}=&\frac{k\cdot k_{l+1}}{(z_2-z_{l+1})^2}\\
a_{n-1}^{l1}=&\frac{k\cdot k_{l+1}}{(z_2-z_{l+1})^2}\\
a_{n-1}^{ll}=&-\sum_i\frac{k_{l+1}\cdot k_i}{(k_{l+1}-z_i)^2}-\frac{k_l\cdot k}{(z_{l+1}-z_2)^2}\\
a_{n-1}^{ij}=&\frac{k_{i+1}\cdot k_{j+1}}{(z_{i+1}-z_{j+1})^2}, \ i\neq j\\
\end{align*}
Or, in matrix form,
\begin{align}
A_{n-1}&=\begin{pmatrix}
-\sum_{i\neq 1,2}\frac{k\cdot k_i}{(z_2-z_i)^2} & \frac{k\cdot k_3}{(z_2-z_3)^2}& ... & \frac{k\cdot k_l}{(z_2-z_l)^2} & ...\\
\frac{k\cdot k_3}{(z_2-z_3)^2} &... & ...\\
... & ... & ...\\
\frac{k\cdot k_l}{(z_2-z_l)^2}&...&...\\
... & ... & ...\\
\frac{k\cdot k_{n-3}}{(z_2-z_{n-3})^2}&...&... \\
\end{pmatrix}	
\end{align}
Keeping  the collinear behavior of the Jacobian determinant in mind, the vertex operator at $z_1$ can be simplified as $z_1 \rightarrow z_2$.
In order to reproduce the right singular behavior at $z_1\rightarrow z_2$ as in eq.(\ref{ja}),  we propose that we can actually integrate out $z_1$ in $\mathcal{O}(z_1)$ as
$\oint_{\hat{z}_1}\frac{e^{k_1\cdot X(z_1)}}{k_1\cdot P(z_1)}\sim \frac{e^{k_1\cdot X(\hat{z}_1)}}{k_1\cdot P'(\hat{z}_1)}$, and the only singular contraction for the denominator part  is
\begin{align*}
 \bra{0}k_1\cdot P(z_1) k_2 \cdot X(z_2)\ket{0}=\frac{k_1\cdot k_2}{z_1-z_2}. 
\end{align*}
 The rest part just takes the limit, $k_1\cdot k_2 \rightarrow 0, \ \ \ \hat{z}_1-\hat{z}_2 \rightarrow 0$. That would produce the right short distance operator product expansion.

One may wonder why $T(w)  \mathcal{O}(\hat{z})$ has nontrivial OPE since by definition T commutes with $ \mathcal{O}$ because $ \mathcal{O}$ corresponds a physical state in ambitwistor string theory. However this is true only if the integrand of $ \mathcal{O}$ is not in the vicinity of $\hat{z}$. Since the integrand of $ \mathcal{O}$ is singular at $\hat{z}$, we expect there is a zero mode insertion at $\hat{z}$, which does not
commute with T.
\begin{align}\label{ve}
 \nonumber	&\oint_{\hat{z_1}}\frac{e^{k_1\cdot X(z_1)}}{k_1\cdot P(z_1)}\left(\epsilon_1\cdot P(z_1)+k_1\cdot\psi(z_1)\epsilon_1\cdot\psi(z_1)\right)J(z_1)dz_1\\  \nonumber
\stackrel{z_1 \rightarrow z_2}{\longrightarrow}& \oint_{\hat{z}_1}\frac{e^{k_1\cdot X(z_1)}\left(\epsilon_1\cdot P(z_1)+k_1\cdot\psi(z_1)\epsilon_1\cdot\psi(z_1)\right)J(z_1)dz_1}{k_1\cdot P(\hat{z}_1)+(z_1-\hat{z}_1)k_1\cdot P^\prime(\hat{z}_1)}\\  
=&\frac{e^{k_1\cdot X(\hat{z}_1)}}{k_1\cdot P'(\hat{z}_1)}\left(\epsilon_1\cdot P(\hat{z}_1)+k_1\cdot\psi(\hat{z}_1)\epsilon_1\cdot\psi(\hat{z}_1)\right)J(\hat{z}_1)
\end{align}
That is to say, this part of $\mathcal{O}_{k_i,\epsilon_i}(z_i)$ has the following OPE,
\begin{align}\label{ja2}
\nonumber &\frac{e^{k_1\cdot X(\hat{z_1})}}{k_1\cdot P'(\hat{z}_1)}\oint_{\hat{z}_2}\frac{e^{k_2\cdot X(z_2)}}{k_2\cdot P(z_2)} dz_2\\  
&=-\frac{(\hat{z_1}-\hat{z}_2)^2}{k_1\cdot k_2}\oint_{\hat{z}_2}\frac{e^{k\cdot X(z_2)}}{k\cdot P(z_2)} dz_2 
\end{align}
with $k=k_1+k_2$, $\oint_{\hat{z}_2}\frac{e^{k\cdot X(z_2)}}{k\cdot P(z_2)}$ is obtained by the following 
product,
\begin{align}
\nonumber &k_1\cdot k_2 \rightarrow 0, \ \ \ \hat{z}_1-\hat{z}_2 \rightarrow 0\\ \nonumber
&	e^{k_1\cdot X(\hat{z}_1)}\oint_{\hat{z}_2}\frac{e^{k_2\cdot X(z_2)}}{k_2\cdot P(z_2)}dz_2\sim \oint_{\hat{z}_2}\frac{e^{k\cdot X(z_2)}}{\frac{k_2\cdot k_1}{z_2-z_1}+\frac{k_2\cdot k_l}{z_2-z_l}} dz_2
     \sim  \oint_{\hat{z}_2}\frac{e^{k\cdot X(z_2)}}{\frac{k_1\cdot k_l}{z_2-z_l}+\frac{k_2\cdot k_l}{z_2-z_l}}dz_2\\
	& =\oint_{\hat{z}_2}\frac{e^{k\cdot X(z_2)}}{\frac{k\cdot k_l}{z_2-z_l}}dz_2=\oint_{\hat{z}_2}\frac{e^{k\cdot X(z_2)}}{k\cdot P(z_2)}dz_2
\end{align}

Next, let us consider the OPE for the numerator part of the vertex operator,
\begin{align}\label{fa}
\nonumber&e^{k_1\cdot X(z_1)}\left(\epsilon_1\cdot P(z_1)+k_1\cdot\psi(z_1)\epsilon_1\cdot\psi(z_1)\right)e^{k_2\cdot X(z_2)}\left(\epsilon_2\cdot P(z_2)+k_2\cdot\psi(z_2)\epsilon_2\cdot\psi(z_2)\right)\\ \nonumber
=&e^{k_1\cdot X(z_1)}\left(\frac{\epsilon_1\cdot k_2}{z_1-z_2}+\frac{\epsilon_1\cdot k_l}{z_1-z_l}+k_1\cdot\psi(z_1)\epsilon_1\cdot\psi(z_1)\right)e^{k_2\cdot X(z_2)}\left(\frac{\epsilon_2\cdot k_1}{z_2-z_1}+\frac{\epsilon_2\cdot k_l}{z_2-z_l}+k_2\cdot\psi(z_2)\epsilon_2\cdot\psi(z_2)\right)\\ \nonumber
\nonumber =&e^{k\cdot X(z_2)}\frac{\epsilon_1\cdot k_2}{z_1-z_2}\frac{\epsilon_2\cdot k_1}{z_2-z_1}+e^{k\cdot X(z_2)}\frac{\epsilon_1\cdot k_2}{z_1-z_2}      
 \left(\frac{\epsilon_2\cdot k_l}{z_2-z_l}+k_2\cdot\psi(z_2)\epsilon_2\cdot\psi(z_2)\right)\\ \nonumber
&+e^{k\cdot X(z_2)}\frac{\epsilon_2\cdot k_1}{z_2-z_1}\left(\frac{\epsilon_1\cdot k_l}{z_1-z_l}+k_1\cdot\psi(z_1)\epsilon_1\cdot\psi(z_1)\right)\\ \nonumber
&+e^{k\cdot X(z_2)}\frac{1}{z_1-z_2}\left((\epsilon_1\cdot k_2)k_1\cdot\psi(z_1)\epsilon_2\cdot\psi(z_2)+(k_1\cdot\epsilon_2)\epsilon_1\cdot\psi(z_1)k_2\cdot\psi(z_2)\right)\\ \nonumber
&-e^{k\cdot X(z_2)}\frac{1}{z_1-z_2}\left((\epsilon_1\cdot\epsilon_2)k_1\cdot\psi(z_1)k_2\cdot\psi(z_2)
+ (k_1\cdot k_2)\epsilon_1\cdot\psi(z_1)\epsilon_2\cdot\psi(z_2)\right)\\ \nonumber
&+e^{k\cdot X(z_2)}\frac{1}{(z_1-z_2)^2}\left((\epsilon_1\cdot k_2) (k_1\cdot\epsilon_2)-(\epsilon_1\cdot \epsilon_2) (k_1\cdot k_2)\right)\nonumber
\end{align}
The double poles cancel each other as desired. The term $\frac{k_1\cdot k_2}{(z_1-z_2)^2}$ is in fact a single pole as we can work out,
\begin{align}
\frac{k_1\cdot k_2}{z_1-z_2} \stackrel{z_1 \rightarrow z_2}{\longrightarrow} \frac{k_2\cdot k_l}{z_2-z_l}=	\frac{1}{2}\frac{(k_2-k_1)\cdot k_l}{z_2-z_l}=	\frac{1}{2}(k_2-k_1)\cdot P(z_2)
\end{align}	
where, summation over $l\neq 1,2$ is assumed. Collecting the remaining terms, we have finally, 
\begin{align}\label{fa2}
	\nonumber&e^{k_1\cdot X(z_1)}\left(\epsilon_1\cdot P(z_1)+k_1\cdot\psi(z_1)\epsilon_1\cdot\psi(z_1)\right)e^{k_2\cdot X(z_2)}\left(\epsilon_2\cdot P(z_2)+k_2\cdot\psi(z_2)\epsilon_2\cdot\psi(z_2)\right)\\ \nonumber
\nonumber =&e^{k\cdot X(z_2)}\frac{1}{z_1-z_2}\left((\epsilon_1\cdot k_2)\epsilon_2-(\epsilon_2\cdot k_1)\epsilon_1+\frac{1}{2}\epsilon_1\cdot \epsilon_2(k_1-k_2)\right)\cdot P(z_2)\\ 
&+e^{k\cdot X(z_2)}\frac{1}{z_1-z_2}k\cdot\psi(z_2)\left((\epsilon_1\cdot k_2)\epsilon_2-(\epsilon_2\cdot k_1)\epsilon_1+\frac{1}{2}\epsilon_1\cdot \epsilon_2(k_1-k_2)\right)\cdot\psi(z_2)
\end{align}
 Combining eq.(\ref{ve}, \ref{ja2}, \ref{fa2}), and the fact that OPE of the Kac-Moody currents just produces a factor of  $1/(z_1-z_2)$, we have,
\begin{align}\label{ope2}
\mathcal{O}_{k_1,\epsilon_1}(z_1)\mathcal{O}_{k_2,\epsilon_2}(z_2)=&	\frac{-1}{k_1\cdot k_2}\sum_{s=\pm} C_{12s}\mathcal{O}_{k,\epsilon^s}(z_2)\\
\label{ope3} C_{12\pm}=&2\left((\epsilon_1\cdot k_2)\epsilon_2-(\epsilon_2\cdot k_1)\epsilon_1+\frac{1}{2}\epsilon_1\cdot \epsilon_2(k_1-k_2)\right)\cdot \epsilon^{\mp} 
\end{align}
The OPE we have check so far, as presented in eq.(\ref{ope2}-\ref{ope3})) is  in  agreement with the result in ref.\cite{Taylor}. This proves that the vertex operator defined in eq.(\ref{ve}) is the right choice at $z_1\rightarrow z_2$.
In what follows we shall keep in mind that $z_1$ is very close to $z_2$, and we shall assume a helicity +1 polarization just for convenience for the vertex operator at $z_1$. Hence we shall work with
\begin{align}\label{oo2}
\mathcal{O}_+(\omega,u,\bar{u},z_1)=\frac{e^{k_1\cdot X(z_1)}}{k_1\cdot P'(z_1)}\left(\epsilon_1^+\cdot P(z_1)+k_1\cdot\psi(z_1)\epsilon_1^+\cdot\psi(z_1)\right)J(z_1)
\end{align}
Viewing the scattering amplitude as the correlation function of the
string vertex insertions, we have to address the following issues
before we discuss the energy momentum tensor insertion in celestial
CFT. First coordinates  are correlated  subjecting to the momentum
conservation, we cannot move one vertex operator freely in
celestial space. Second, coordinates are further restricted by the
scattering equation, eq.(\ref{se}), so we have to find a solution to  eq.(\ref{se}),
in which  one coordinate is a free variable. To solve those issues,
we resort to the one complex parameter continuation of the momentum $k$ as the same technique used
in BCFW method.
\begin{align}\label{cs}
k_1&\rightarrow k_1+u \tilde{\lambda}_1\lambda_n\\
k_n&\rightarrow k_n-u\tilde{\lambda}_1\lambda_n  ,
\end{align}
so that the momentum conservation is not violated and we get one free parameter $u$. In the following
we shall  consider how to construct a stress tensor in celestial CFT
with correct OPE to the vertex operator sitting in the coordinate $z_1$.
Using $SL(2)$ invariance, we can fix the n'th coordinate to
infinity, so that we are left with the following scattering equation
to solve with,
\begin{equation}\label{se2}
\sum_{j\neq i,n} \frac{k_i\cdot k_j}{z_i-z_j}=0, \ \ i=1,2,...,n-3.
\end{equation}
Next, we consider the collinear limit, $k_1\cdot k_2\rightarrow 0$,
we seek for the solution with the behavior  $z_1\rightarrow z_2$,
such that
\begin{equation}\label{z1}
\sum_{j\neq i,1,n} \frac{k_2\cdot k_j}{z_2-z_j}=\frac{k_1\cdot
k_2}{z_1-z_2}
\end{equation}
is finite.

Under these circumstances, we have the solution eq.(\ref{z1}) for $z_1$ to eq.(\ref{se2}) with
$z_1-z_2$ varying linear in $k_1\cdot k_2$, while keeping fixed
$z_i, i=2,3,...,n-1$, which are determined according to the remaining equations in eq.(\ref{se2}),
\begin{align}
&\sum_{j\neq 1,2,n} \frac{k\cdot k_j}{z_2-z_j}=0,\\
&\frac{k_i\cdot k}{z_i-z_2}+\sum_{j\neq i,1,2,n} \frac{k_i\cdot k_j}{z_i-z_j}=0, \ \ i=3,...,n-3.
\end{align}
with $k=k_1+k_2$. That is, the above equations are just the scattering equations for n-1 particles with
$k_2$ replaced by $k$.

For fixed $z_i, i=2,3,...,n-1$, We solve the  remaining equation for $z_1$.
\begin{align}\label{li}
z_1-z_2=\frac{k_1\cdot k_2}{\sum_{j\neq 1,2,n} \frac{k_2\cdot
		k_j}{z_2-z_j}}
\end{align}	
Equation (\ref{li}) provides a link between the two coordinate systems, one in
string world sheet and another in celestial sphere, respectively. To make
things more transparent, let us decompose the momentum into the
following form,  $k=\omega q$,
$\omega=k^0-k^3$,
$q=\frac{k}{k^0-k^3}$. Defining $u=\frac{k^1+i
k^2}{k^0-k^3},\ \bar{u}=\frac{k^1-i k^2}{k^0-k^3} $, we can
represent the momentum $q$ in a matrix form,
\begin{align*}
q=&q^{\mu}\bar{\sigma}_\mu=\frac{1}{k^0-k^3}\begin{pmatrix}k^0+k^3 & k^1-i k^2\\ k^1+i k^2 & k^0-k^3\end{pmatrix}\\
=&\begin{pmatrix}u\bar{u} & \bar{u}\\ u &1\end{pmatrix}=\begin{pmatrix}\bar{u} \\ 1\end{pmatrix} \begin{pmatrix}u &1 \end{pmatrix}=\bar{\xi}\xi\\
\bar{\xi}=&\begin{pmatrix}\bar{u} \\ 1\end{pmatrix}, \ \ \xi=\begin{pmatrix}u & 1\end{pmatrix}\\
q^\mu=&\frac{1}{2}\begin{pmatrix}u\bar{u}+1,& u+\bar{u},& -i(u-\bar{u}),&u\bar{u}-1\end{pmatrix}=\frac{1}{2}\xi \sigma^\mu \bar{\xi}
\end{align*}
We recognize $u$ and $\bar{u}$ are the coordinates in
celestial sphere, and $\omega$ the lightcone momentum. we have
\begin{equation}
k_1\cdot k_2\rightarrow (k_1+v\tilde{\lambda}_1\lambda_n)\cdot k_2
\end{equation}

 we introduce the following null 4-vectors in the matrix form,
\begin{align*}
\epsilon^+&=\partial_u q=\begin{pmatrix} \bar{u} & 0\\ 1 & 0\end{pmatrix}=\begin{pmatrix}\bar{u} \\ 1\end{pmatrix} \begin{pmatrix}1& 0\end{pmatrix}=\bar{\xi}\rho\\
\epsilon^-&=\partial_{\bar{u} }q=\begin{pmatrix}u & 1\\ 0 & 0\end{pmatrix}=\begin{pmatrix}1\\ 0\end{pmatrix} \begin{pmatrix}u& 1\end{pmatrix} =\bar{\rho}\xi\\
\mu &=\partial_{\bar{u}} \partial_u q=\begin{pmatrix}1 & 0\\ 0 & 0\end{pmatrix}=\begin{pmatrix}1\\ 0\end{pmatrix} \begin{pmatrix}1& 0\end{pmatrix}=\bar{\rho}\rho\\
\end{align*}
The only non-vanishing inner products between those four null
4-vectors are the following,
\begin{align*}
q\cdot \mu&=-\frac{1}{2}\\
\epsilon^+\cdot\epsilon^-&=\frac{1}{2}
\end{align*}
 it is convenient to introduce the following orthogonal basis for the $X$ and $P$ fields,
\begin{align*}
P^{(1)}(z)&=q\cdot P(z)\\
X^{(1)}(z)&=\mu\cdot X(z)\\
P^{(2)}(z)&=\mu\cdot P(z)\\
X^{(2)}(z)&=q\cdot X(z)\\
P^{(3)}(z)&=\epsilon^{+}\cdot P(z)\\
X^{(3)}(z)&=\epsilon^{-}\cdot X(z)\\
P^{(4)}(z)&=\epsilon^{-}\cdot P(z)\\
X^{(4)}(z)&=\epsilon^{+}\cdot X(z)\\
\end{align*}
In order that the BCFW momentum shift defined in eq.(\ref{cs}) transforms into
the celestial coordinate $u_1$ shift, it's convenient to fix
$k_n\sim \mu$ using Lorentz invariance,
\begin{equation}
k_n=k_n^0\begin{pmatrix}1 & 0\\ 0 &
0\end{pmatrix}=k_n^0\begin{pmatrix}1\\ 0
\end{pmatrix}\begin{pmatrix}1 & 0\end{pmatrix}
\end{equation}
We have
\begin{align*}
\delta k_1&=-\delta k_n=u\omega_1\epsilon^+_1=u\omega_1 \begin{pmatrix}\bar{u}_1\\1 \end{pmatrix}\begin{pmatrix} 1&0\end{pmatrix}\\
k_1&\rightarrow k_1+u\omega_1\epsilon^+_1=\omega_1\begin{pmatrix}\bar{u}_1\\1 \end{pmatrix}\begin{pmatrix} u_1+u&1\end{pmatrix}\\
k_n&\rightarrow k_1-u\omega_1\epsilon^+_1=\begin{pmatrix}k_n^0-u\omega_1\bar{u}_1\\-u\omega_1 \end{pmatrix}\begin{pmatrix} 1& 0\end{pmatrix}\\
k_1&\cdot k_2\rightarrow (k_1+u\omega_1\epsilon^+_1)\cdot
k_2=-\frac{\omega_1\omega_2}{2}(u+u_1-u_2)(\bar{u}_1-\bar{u}_2)
\end{align*}

\section{The  Energy Momentum Tensor in collinear limit}\label{emt}
The correlation function in celestial space is related  to the
scattering amplitude by the inverse Mellin transformation
\begin{equation}
A_{s_1...s_n}(|\omega|_1,u_1,\bar{u}_1,...,|\omega|_n,u_n,\bar{u}_n)=(\prod_i
\int d\Delta_i
|\omega|_i^{-\Delta_i})\mathcal{A}_{s_1...s_n}(|\Delta|_1,u_1,\bar{u}_1,...,|\Delta|_n,u_n,\bar{u}_n)
\end{equation}
where, $\Delta_i=h_i+\bar{h}_i$ is the dilation weight with $h_i$
$(\bar{h}_i$) the conformal weight for the left(right)-moving part,
$S_i=h_i-\bar{h}_i$ is the spin, $s_i$ the polarization helicity.  So for each celestial
 primary fields with definite spin $S$, we have $h=\frac{\Delta+S}{2}$, and we expect an energy momentum tensor with the following OPE
 \begin{equation}\label{to1}
  \mathcal{T}(v) \mathcal{O}_s(\Delta, u_1,\bar{u}_1)=\frac{1}{v-u_1}\partial_{u_1}  \mathcal{O}_s(\Delta, u_1,\bar{u}_1)+\frac{(\Delta+S)/2}{(v-u_1)^2} \mathcal{O}_s(\Delta, u_1,\bar{u}_1)
 \end{equation}
 Define
 \begin{align}\label{ct}
 C=\frac{u+u_1-u_2}{z_1-z_2}&=\frac{2}{\omega_1\omega_2(\bar{u}_1-\bar{u}_2)}\sum_{j\neq i,1,n} \frac{k_2\cdot k_j}{z_2-z_j}\\
\label{ct1} &=-\frac{2}{\omega_2(\bar{u}_1-\bar{u}_2)}\sum_{j\neq i,1,n} {\frac{q_1\cdot k_j}{z_2-z_j}} \Biggr|_{u_1=u_2}
  \end{align}
  In eq.(\ref{ct}-\ref{ct1}), the singular behavior is holomorphic, we retain this limit for the helicity +1  polarization. Similarly, for helicity -1 polarization, we would choose anti-holomorphic OPE expansion. So the following OPE works for  $\mathcal{O}_{+}$,
  and we have $\partial_u \epsilon^+=0$. We recognize that in the collinear limit, the primary field in CCFT, $\mathcal{O}_s(\Delta, u_1,\bar{u}_1)$ defined in eq.(\ref{to1}), is the Mellin transformed  vertex operator in ambitwistor string theory,  $\mathcal{O}_+(\omega, u_1,\bar{u}_1,z_1)$ in eq.(\ref{oo2}), and $u_1$ and $z_1$ are related by a global conformal transformation, eq.(\ref{ct} or \ref{ct1}). Hence we have the following OPE
  \begin{align}\label{to2}
  T(w)&=(\frac{dv}{dw})^2\mathcal{T}(v) =C^2\mathcal{T}(v)  \\  \label{to3} 
 T(w)&\mathcal{O}_+(\omega, u_1,\bar{u}_1,z_1)=\left(\frac{1}{w-z_1}(\partial_{z_1} +C\partial_{u_1}  )+\frac{(1-\omega\partial_{\omega})}{2(w-z_1)^2}\right)\mathcal{O}_+(\omega, u_1,\bar{u}_1,z_1)
  \end{align}
 In order the OPE, eq.(\ref{to3}), be satisfied, we propose the following form of the  energy-momentum tensor, which is equivalent, by a global conformal transformation, to the stress tensor in CCFT in the collinear limit,
\begin{align}\label{emt1}
T(z)
\nonumber =&\partial_z X(z)\cdot P(z)+T_{g}+T_{gh}\\ \nonumber
&+\frac{1}{2} C\left((X_1(z)-i X_2(z)) (P^0(z)-P^3(z))+(X_0(z)+X_3(z)) (P^1(z)-i P^2(z))\right)\\ 
&+\frac{1}{2}\partial_z\left(X(z)\cdot P(z)\right)\\ 
\label{cc1}c_{total}=&(11+\frac{1}{2})d-52+11+c_g
\end{align}
Here, $d$ is the bulk flat spacetime dimension, $c_{gh}$ is for the two pair of fermionic $(2,-1)$ ghosts and one pair of $(3/2, -1/2)$ bosonic ghosts, $c_g$ is the center charge for the Sugawara construction of the stress tensor for the level $k$  WZNW model.  In what follows we are going to check if the stress tensor defined in eq.(\ref{emt1}-\ref{cc1}) satisfies the correct OPE. 

With the energy momentum tensor defined as in eq.(\ref{emt1}),  the OPE between $T(w)$ and the vertex operator at $z_1$, defined in eq.(\ref{oo2}) can be checked.

\begin{align}\label{to}
\nonumber &T(w)\mathcal{O}_+(\omega,u,\bar{u},z_1)\\ \nonumber
=&T(w)\frac{e^{k\cdot X(z_1)}}{k\cdot P'(z_1)}(\epsilon^+\cdot P(z_1)+k\cdot\psi(z_1)\epsilon^+\cdot\psi(z_1))J(z_1)\\ \nonumber
=&\frac{1}{w-z_1}\partial_{z_1} \mathcal{O}_+(\omega,u,\bar{u},z_1)\\ \nonumber
&+\frac{1}{2}\frac{C}{w-z_1}\left((X_1(z_1)-i X_2(z_1)) (k^0-k^3)+(X_0(z_1)+X_3(z_1)) (k^1-i k^2)\right) \mathcal{O}_+(\omega,u,\bar{u},z_1)\\ \nonumber
&+\frac{1}{2}\frac{C}{(w-z_1)^2}\left((k_1-i k_2) (P^0(w)-P^3(w))+(k_0+k_3) (P^1(w)-i P^2(w))\right)\frac{\mathcal{O}_+(\omega,u,\bar{u},z_1)}{k\cdot P'(z_1)}\\ \nonumber
&+\frac{1}{2}\partial_w\left(\frac{k\cdot X}{w-z_1}\mathcal{O}_+(\omega,u,\bar{u},z_1)-\frac{\epsilon^+\cdot P(z_1)}{w-z_1}\frac{e^{k\cdot X(z_1)}}{k\cdot P'(z_1)}J(z_1)+\frac{k\cdot P(w)}{(w-z)^2k\cdot P'(z_1)}\mathcal{O}_+(\omega,u,\bar{u},z_1)\right)\\ \nonumber
=&\frac{1}{w-z_1}\partial_{z_1} \mathcal{O}_+(\omega,u,\bar{u},z_1)\\ \nonumber
&+\frac{C}{w-z_1}(\partial_u k \cdot X(z_1))\mathcal{O}_+(\omega,u,\bar{u},z_1)\\ \nonumber
&-\frac{C}{(w-z_1)^2}\left(\partial_u k\cdot P(w)\right)\frac{\mathcal{O}_+(\omega,u,\bar{u},z_1)}{k\cdot P'(z_1)}\\  
&-\frac{1}{2(w-z)^2}\left(k\cdot X \mathcal{O}_+(\omega,u,\bar{u},z_1)-\epsilon^+\cdot P(z_1)\frac{e^{k\cdot X(z_1)}}{k\cdot P'(z_1)}J(z_1)+\mathcal{O}_+(\omega,u,\bar{u},z_1)\right) 
\end{align}
We now expand one of the double pole terms in eq.(\ref{to}),
\begin{align}
\nonumber	&-\frac{C}{(w-z_1)^2}\left(\partial_u k\cdot P(w)\right)\frac{\mathcal{O}_+(\omega,u,\bar{u},z_1)}{k\cdot P'(z_1)}\\ \nonumber
	=&-C\left(\frac{\partial_u k\cdot P(z_1)}{(w-z_1)^2}+\frac{\partial_u k\cdot P'(z_1)}{w-z_1}\right)\frac{\mathcal{O}_+(\omega,u,\bar{u},z_1)}{k\cdot P'(z_1)}\\
	=&-\left(-\frac{ k\cdot P'(z_1)}{(w-z_1)^2}+C\frac{\partial_u k\cdot P'(z_1)}{w-z_1}\right)\frac{\mathcal{O}_+(\omega,u,\bar{u},z_1)}{k\cdot P'(z_1)} 
\end{align}
Here, we used the following relation. From $\frac{ d }{du}(k\cdot P(z))=\partial_u k\cdot P(z)+k\cdot \frac{1}{C}\partial_z P(z)=0$, we have $C \partial_u k\cdot P(z_1)=- k\cdot P'(z_1)$.
Collecting terms with single pole and double pole separately, we have
\begin{align}
\nonumber &T(w) \mathcal{O}_+(\omega,u,\bar{u},z_1)\\ \nonumber
=&\frac{1}{w-z_1}\partial_{z_1}\mathcal{O}_+(\omega,u,\bar{u},z_1)+\frac{C}{w-z_1}\left(\partial_u k \cdot X(z)-\frac{\partial_u k\cdot P'(z)}{k\cdot P'(z)} \right)\mathcal{O}_+(\omega,u,\bar{u},z_1)\\ \nonumber
&+\frac{1}{(w-z_1)^2}\left((\frac{1}{2}-\frac{k}{2}\cdot X(z))\mathcal{O}_+(\omega,u,\bar{u},z_1)+\frac{1}{2}\epsilon^+\cdot P(z)\frac{e^{k\cdot X(z)}}{k\cdot P'(z)}J(z_1)\right)\\ \nonumber
=&\left(\frac{1}{w-z_1}(\partial_{z_1}+C\partial_u)+\frac{1}{(w-z_1)^2}\frac{1}{2}(1-\omega\partial_\omega)\right)\mathcal{O}_+(\omega,u,\bar{u},z_1)\\
=&\left(\frac{1}{w-z_1}(\partial_{z_1}+C\partial_u)+\frac{1}{(w-z_1)^2}\frac{1}{2}(1-\omega\partial_\omega)\right)\frac{e^{k\cdot X(z_1)}}{k\cdot P'(z_1)}(\epsilon^+\cdot P(z_1)+k\cdot\psi(z_1)\epsilon^+\cdot\psi(z_1))J(z_1) 
\end{align}
So the energy momentum tensor defined in eq.(\ref{emt1}) has the right OPE form, eq.(\ref{to2}),  with respect to the vertex operator $\mathcal{O}(\omega,u,\bar{u},z_1)$ defined in eq.(\ref{oo2}). Thus, $T(w)$ and $\mathcal{O}(\omega,u,\bar{u},z_1)$, originally defined on the 2d worldsheet, can be  extended to the celestial sphere via a global conformal transformation defined by eq.(\ref{ct} or \ref{ct1}).

Calculating the center charge for CCFT for the Yang-Mills theory, we first write the energy momentum tensor in a more compact form.
\begin{align}
T(w)=&T^{(0)}(w)+I(w)+\partial_w J(w) \\
T^{(0)}(w)=& \partial_z X(w)\cdot P(w)+T_{g}(w)+T_{gh}(w)\\
I(w)=&	\frac{1}{2} C\left((X_1(w)-i X_2(w)) (P^0(w)-P^3(w))+(X_0(w)+X_3(w)) (P^1(w)-i P^2(w))\right)\\
J(w)=&\frac{1}{2}X(w)\cdot P(w)
\end{align}
We have,
\begin{align}
	T^{(0)}(w) I(z)=&\frac{I(w)}{(w-z)^2}\\
	T^{(0)}(w) J(z)=&\frac{J(w)}{(w-z)^2}+\frac{d}{2(w-z)^3}\\
	I(w) I(z)=&0\\
	I(w)J(z)=&0\\
	J(w) J(z)=&-\frac{d}{4(w-z)^2}
\end{align}
This leads to
\begin{align}
\nonumber T(w)T(z)=&(T^{(0)}(w)+I(w)+\partial_w J(w) )) (T^{(0)}(z)+I(z)+\partial_z J(z) ))\\ \nonumber
&=\frac{\partial_zT^{(0)}(z)}{w-z}+\frac{2T^{(0)}(z)}{(w-z)^2}+\frac{2d+d/2+c_g-52+11}{2(w-z)^4}\\ \nonumber
&+\frac{I(w)}{(w-z)^2}+\frac{2J(w)}{(w-z)^3}+\frac{3d}{2(w-z)^4}\\ \nonumber
&+\frac{I(z)}{(w-z)^2}+\frac{2J(z)}{(z-w)^3}+\frac{3d}{2(w-z)^4}+\frac{3d}{2(w-z)^4}\\  
&=\frac{\partial_zT(z)}{w-z}+\frac{2T(z)}{(w-z)^2}+\frac{11d+d/2+c_g-41}{2(w-z)^4} \\
\label{cc} c_{total}=&(11+\frac{1}{2})d-41+c_g 
\end{align}
Hence we have finished the proof that eq.(\ref{emt1}-\ref{cc1}) are indeed the right choice for the primary field $\mathcal{O}(\omega,u,\bar{u},z_1)$ defined in eq.(\ref{oo2}).

\section{Finite Size Effect}\label{fse}
So far we have considered CCFT as a 2d CFT living on the celestial sphere at null infinity. However since our universe has finite 
size both in space and time, we need to consider the finite size effect as a perturbative non-CFT correction to the CCFT we have considered so far. That is to say, CCFT is living very close to but not exactly at the boundary of the $AdS_3$ slices.  Similar situation has been considered by the authors of of ref.\cite{verlinde}, who proposed that the removal of the asymptotic region  beyond a  radial distance $r_c$ in the  $AdS_3$ bulk theory is equivalent to defining a quantum field theory on a Dirichlet wall at  the  radial distance $r=r_c$. That quantum field theory can be obtained by adding a $T\bar{T}$ term as perturbation to the original 2d CFT when $r_c$ is large. In particular, the vacuum to vacuum amplitude is modified by an exponential term, $\bra{0}\exp\left(i\mu\int dz d\bar{z}T(z')\bar{T}(\bar{z})\right)\ket{0}$. and $\mu \sim \frac{1}{r_c}$. 
Since $T\bar{T}$ perturbed CFT is a big subject following  the original refs.\cite{Zamolodchikov0401146, Zamolodchikov1608.05499}, we stop short of going into details
on the development of the $T\bar{T}$ perturbed CFT. Here we just make use of  the main result in ref\cite{verlinde} by proposing that
the perturbation parameter for the CCFT should be $\mu \sim \frac{1}{t_u}$. Here $t_u \sim r_c \sim 8\cdot 10^{60}$ is the expansion age of the present universe in Planck units. We make this choice for $\mu$ since $t_u$ is the only scale available in CCFT besides the Planck scale. 
Adding this new term, we are ready to calculate its effect on the vacuum to vacuum amplitude,
\begin{align}
	A&=\bra{0}\exp\left(\frac{i}{t_u}\int dz d\bar{z}T(z')\bar{T}(\bar{z})\right)\ket{0}\\
	&\sim 1-\bra{0}\frac{1}{2t_u^2}\int dz d\bar{z}T(z)\bar{T}(\bar{z})\int dz' d\bar{z}'T(z')\bar{T}(\bar{z}')\ket{0}\\
	&=1-\frac{1}{2t_u^2}\oint_{z'} dz\int_{-\frac{1}{\epsilon}}^{\frac{1}{\epsilon}} dz' \frac{c}{2(z-z')^4}\oint_{\bar{z}'} d\bar{z} \int_{-\frac{1}{\epsilon}}^{\frac{1}{\epsilon}}d\bar{z}'\frac{\bar{c}}{2(\bar{z}-\bar{z}')^4} 
\end{align}

This integral needs to be regularized since it is of the type $0\cdot \infty$. We just put an $\epsilon$ regularization,

\begin{align}
	A&=1-\frac{c\bar{c}}{2t_u^2}\lim_{\epsilon \rightarrow 0}
	\oint_{z'}dz\frac{ (z-z'+1)^\epsilon}{2(z-z')^4}
	\int_{-\frac{1}{\epsilon}}^{\frac{1}{\epsilon}}dz'
	\oint_{\bar{z}'}d\bar{z}\frac{(\bar{z}-\bar{z}'+1)^\epsilon}{2(\bar{z}-\bar{z}')^4}
	\int_{-\frac{1}{\epsilon}}^{\frac{1}{\epsilon}}d\bar{z}'\\
	&=1-\frac{c\bar{c}}{2t_u^2}\lim_{\epsilon \rightarrow 0}
	\epsilon(\epsilon-1)(\epsilon-2)\frac{1} {2\cdot 6}
	\frac{2}{\epsilon}
	\epsilon(\epsilon-1)(\epsilon-2)\frac{1} {2\cdot 6}
	\frac{2}{\epsilon}\\
\label{lambda}	&\sim \exp(-\frac{c\bar{c}}{18}\Lambda)
\end{align}
Here, $c=\bar{c}=11d+d/2+c_g-41$ for the Yang-Mills theory as in eq.(\ref{cc}) and $\Lambda\sim \frac{1}{t_u^2}\sim 10^{-122} $ \cite{shaw} is the observed value for the present cosmological constant in Planck units. The integral we encounter here is very much like the one appeared in ref.\cite{Cardy1986}, but the final form of the finite size effects differs. It is very tentative to say that our result generalizes the finite length effect in 2d CFT to the finite surface effect in $AdS_3$.
We immediately recognize that the finite size effect in CCFT takes the same form as the cosmological term in 4d gravitational theories.
This should be of no surprise since both are kinds of vacuum energy of the Casimir type, and certainly related to the length scale of our universe. Still,  this feature does not manifest itself had we not worked on the holographic dual to the 4d gauge theories. In fact, much effort has been made in understanding the nature of the tiny cosmological constant being observed \cite{weinberg} and  in understanding the nature of the dark energy contributed to the evolution of our universe. Here we simply argue that any quantum field theory in 4d spacetime  be modified by the finite size effect of the type as in eq(\ref{lambda}), should this quantum field theory have a dual description in terms of 2d CCFT on celestial sphere at the null infinity. It is also possible that the cosmological constant is not really a ``constant" but actually evolves with time, as the length scale of our universe does.

\section{Conclusion}\label{c}
The development in CCFT is fascinating since it is a holographic dual description to the 4d quantum field theories  in Minkowski spacetime. Being a dual description, CCFT accommodates an infinite-dimensional asymptotic symmetry, BMS symmetry group. In fact, BMS symmetry as seen in 4d quantum field theory is just the Virasoro and Kac-Moody type of symmetry algebra as seen in many 2d CFT's. Nevertheless, CCFT is not a rational CFT of any kind we know of, neither there exists an operator formalism in the form of free field realization nor known integrable models. Thus the BMS symmetry algebra discussed in the literature are largely classical or semi-classical. The present paper is a modest step taken towards the quantum realization of the BMS symmetry in CCFT, which is a dual description to the 4d quantum filed theories. The benefit of a quantum realization of the BMS symmetry algebra is obvious since it reveals some quantum mechanical effects not seen directly in the 4d realization  of the quantum field theories in Minkowski spacetime. A concrete example is the center charge and the related finite size effect in CCFT, which we managed to calculate in the present paper. We argue that the finite size effect in CCFT is the dual description to the cosmological term in 4d bulk theories. 

 In calculating the finite size effect, we adapted the method used in \cite{verlinde} without explicit proving their result. This procedure can be refined if we start with the warped $AdS_3$ space, and consider how the finite size is incorporated into the Witten diagram \cite{Witten9802150}. This is in the plan of our future work.

\section{Note Added}\label{na}
We are aware that the holomorphic currents in CCFT are intimately related to the soft theorems in 4d  gravity and gauge theories. It is much expected that the center extension can be read off  from the double soft limits. However, several groups have claimed that the center extension thus obtained is zero. So we have to clarify why we get potentially non-zero results here. 

From eq.(\ref{cc}), we get $c_{total}=(11+\frac{1}{2})d-41+c_g$. Then BRST
invariance in ambitwistor string theory requires $c_{s}=(2+\frac{1}{2})d-41+c_g=0$. Hence we have $c_{total}=9d$. For $d=4$. we have $c_{total}=36$. The apparent discrepancy between our result and the others’ can be explained as follows.

First, let us pay attention to the Kac-Moody currents which are related to the zeroth order of the soft theorem in gauge theories. From CHY formalism, the short distance operator product expansion between the Kac-Moody currents are always singular with single poles, so we do not expect to get double pole when two external lines go soft. Double pole arises only for the total contraction between two external lines with the conjugate color indices and opposite helicities. So it is sufficient to consider only the single soft limit. Let us  send  $z$, the coordinate on the celestial sphere for the soft particle, to infinity. Around infinity, $J(z)$ can be expanded as $\bra{0}\sum_{n\in N^+} J_n z^{-n-1}\sim \bra{0}J_1 z^{-2}$. See eq.(4.4-4.5) in ref.\cite{Strominger1503.02663}. The leading $1/z^2$ behavior is an indication that the level of the Kac-Moody algebra $k$ is not zero, since the term $\bra{0}J_1 z^{-2}$ is going to be contracted with the term proportional to $J_{-1}\ket{0} $ leading to $\bra{0}J_1 z^{-2} J_{-1}\ket{0} =\frac{k}{2z^2} $. Here, $k$ is the level of the Kac-Moody algebra. The large $z$ behavior can be checked explicitly for  the single soft limit of the three and four point function \cite{Strominger1706.03917}. See also \cite{Mason1905.09224}. 

Second, similar argument applies to the gravity scattering amplitude. If we consider the n-graviton scattering amplitude with one graviton at $z$ going soft, we indeed find the correct $1/z^4$  behavior for $z$ large when the shadow form of the stress tensor is inserted at $z$. See eq.(4.6-4.9) in ref.\cite{Strominger1609.00282}. This $1/z^4$ behavior precisely implies that 
 $\bra{0}L_2 z^{-4} L_{-2}\ket{0} =\frac{c}{2z^4} $ is non-vanishing. Here c is the center charge for the stress tensor inserted.

Third, there is another possibility that the center charge $c$ for the stress tensor $T(z)$ is zero but $\bra{0}L_2 z^{-4}$ is not paired with  the state $L_{-2}\ket{0}$. Instead, there  may exist another spin 2 holomorphic current
$t(z)$ such that $T(z) t(0)\sim \frac{b}{z^4}$, and  $\bra{0}L_2 z^{-4} l_{-2}\ket{0} =\frac{b}{z^4} $. This possibility leads to a logarithmic CCFT proposed in ref.\cite{Ruzziconi}. At the moment, we can not conclude if CCFT is logarithmic or not, but our construction of the stress tensor seems to suggest a non-zero value  of $c$ for the total energy momentum tensor in the corresponding CCFT.

Fourth, $T\bar{T}$ perturbed CCFT may play an important role in regularizing the loop integrals in the bulk theory providing a UV-completeness, see ref.\cite{Song-He}. But here in our consideration, the  $T\bar{T}$ perturbation  is relevant even at the tree level computation of the scattering amplitude. Of course, since the value of the center charge is closely related to the vacuum energy of the quantum field theories, it is desirable to find its 4d origin which may make 4d-2d holography dictionary more complete.   
  
\acknowledgments
I am grateful to Yihong Gao, Song He, Gang Yang, Ronggen Cai, Chi Zhang, Yaozhong Zhang for useful discussions at the various stage of the present project.  I am indebted to Jens Lyng Petersen for carefully reading the manuscript and for valuable communications.  This work is supported in part by the National Natural Science Foundation of China Grants No.11675240.


\begin{thebibliography}{99}
  \bibitem{bms} 
  H.~Bondi, M. G. J.~van der Burg and A. W. K.~Metzner, \emph{Gravitational waves in general relativity VII.
  	Waves from isolated axisymmetric systems}, {\emph{Proc. Roy. Soc. Lond.}  {\bfseries A 269} (1962) 21};
  	R. K.~Sachs, \emph{Gravitational waves in general relativity VIII. Waves in asymptotically flat space-time},
  	{\emph{Proc. Roy. Soc. Lond.} {\bfseries A 270}  (1962) 103}.
  	
	\bibitem{bt} 
	G.~Barnich and C.~Troessaert,
    \emph{Symmetries of asymptotically flat 4 dimensional spacetimes at null infinity revisited},
	 {\emph{Phys.\ Rev.\ Lett.}\  {\bfseries 105} (2010) 111103};
	 [\href{https://arxiv.org/abs/arXiv:0909.2617}{{\ttfamily
	 		arXiv:0909.2617} [gr-qc]}]; \emph{Supertranslations call for superrotations},
 		{\emph{PoS CNCFG} {\bfseries 2010} (2010) 010}; 
	 [\href{https://arxiv.org/abs/arXiv:1102.4632}{{\ttfamily
	 		arXiv:1102.4632} [gr-qc]}];  \emph{BMS charge algebra},
 \href{https://doi.org/10.1007/JHEP12(2011)105} {\emph{Journal of High Energy Physics} {\bfseries 2011} (2011)105};
[\href{https://arxiv.org/abs/arXiv:1106.0213}{{\ttfamily
		arXiv:0909.2617} [hep-th]}].
	
	\bibitem{banks} 
	T.~Banks,
    \emph{A Critique of pure string theory: Heterodox opinions of diverse dimensions}, 
	[\href{https://arxiv.org/abs/arXiv:hep-th/0306074}{{\ttfamily
			arXiv:hep-th/0306074}}].
		
	\bibitem{stronminger1308.0589} 
	 A.~Strominger, \emph{Asymptotic symmetries of Yang-Mills theory}, [\href{https://doi.org/10.1007/JHEP07(2014)151}
	 {\emph{Journal of High Energy Physics} {\bfseries 2014} (2014) 151},
	[\href{https://arxiv.org/abs/arXiv:1308.0589}{{\ttfamily arXiv:1308.0589}}].
		
		\bibitem{Strominger1406.3312} 
		Daniel Kapec, Vyacheslav Lysov, Sabrina Pasterski and  Andrew Strominger, \emph{Semiclassical Virasoro Symmetry of the Quantum Gravity S-Matrix}, [\href{https://doi.org/10.1007/JHEP08(2014)058}
		{\emph{Journal of High Energy Physics} {\bfseries 1408} (2014) 058},
		[\href{https://doi.org/10.48550/arXiv.1406.3312}{{\ttfamily arXiv:1406.3312}}].	
		
		\bibitem{Strominger1503.02663} 
		Temple He, Prahar Mitra and  Andrew Strominger, \emph{2D Kac-Moody Symmetry of 4D Yang-Mills Theory}, [\href{https://doi.org/10.1007/JHEP10(2016)137}
		{\emph{Journal of High Energy Physics} {\bfseries 2016} (2016)  137},
		[\href{https://doi.org/10.48550/arXiv.1503.02663}{{\ttfamily arXiv:1503.02663}}].	
		
		
		\bibitem{Strominger1609.00282} 
		Daniel Kapec, Prahar Mitra, Ana-Maria Raclariu, Andrew Strominger, \emph{A 2D Stress Tensor for 4D Gravity},  {\emph{Phys. Rev. Lett.} {\bfseries 119}(2017)
		121601}, 
		[\href{https://doi.org/10.48550/arXiv.1609.00282}{{\ttfamily arXiv:1609.00282}}].	
		
		
	\bibitem{cachazo2014scattering}
		F.~Cachazo, S.~He and E.~Y. Yuan, \emph{Scattering of massless particles in
			arbitrary dimensions}, {\emph{Physical review letters} {\bfseries 113} (2014)
			171601}, [\href{https://arxiv.org/abs/arXiv:1307.2199}{{\ttfamily
				arXiv:1307.2199}}].
	
	
	\bibitem{cachazo2013scattering}
	F.~Cachazo, S.~He and E.~Y. Yuan, \emph{Scattering of massless particles:
		scalars, gluons and gravitons},
	\href{http://dx.doi.org/10.1007/JHEP07(2014)033}{\emph{Journal of High Energy
			Physics} {\bfseries 2014} (2014) 33},
	[\href{https://arxiv.org/abs/arXiv:1309.0885}{{\ttfamily arXiv:1309.0885}}].
	
		\bibitem{mason2013ambitwistor}
		L.~Mason and D.~Skinner, \emph{Ambitwistor strings and the scattering
			equations}, \href{http://dx.doi.org/10.1007/JHEP07(2014)048}{\emph{Journal of
				High Energy Physics} {\bfseries 2014} (2014) 48},
		[\href{https://arxiv.org/abs/arXiv:1311.2564}{{\ttfamily arXiv:1311.2564}}].
	
	\bibitem{seiberg9806194}
		A.~Giveon, D.~Kutasov and  N.~Seiberg,  \emph{Comments on String Theory on $AdS_3$}, {\emph{Adv.Theor.Math.Phys.} {\bfseries 2} (1998)
			733-780}, [\href{https://doi.org/10.48550/arXiv.hep-th/9806194}{{\ttfamily
				arXiv:hep-th/9806194}}].
			
	\bibitem{yum9812216}		
       Ming Yu and Bo Zhang, \emph{Light-Cone Gauge Quantization of String Theories on $AdS_3$ Space}, {\emph{Nucl.Phys.} {\bfseries B551} (1999)
       425-449}, [\href{https://doi.org/10.48550/arXiv.hep-th/9812216}{{\ttfamily
       		arXiv:hep-th/9812216}}].			
			
			\bibitem{Frolov}		
			V. P. Frolov, \emph{Null Surface Quantization and Quantum Field Theory in Asymptotically Flat Space-Time}, {\emph{Fortschritte der Physik} {\bfseries 26-9} (1978)
				455-500}.
			
			\bibitem{Strominger1701.00049} 
			Sabrina Pasterski, Shu-Heng Shao and Andrew Strominger, \emph{Flat Space Amplitudes and Conformal Symmetry of the Celestial Sphere},
			{\emph{Phys. Rev.} {\bfseries D 96,} ((2017)  065026},
			[\href{https://doi.org/10.48550/arXiv.1701.00049}{\ttfamily arXiv:1701.00049}].	
			
			
			\bibitem{Strominger1706.03917} 
			Sabrina Pasterski, Shu-Heng Shao, Andrew Strominger, \emph{Gluon Amplitudes as 2d Conformal Correlators}, 
			{\emph{Phys. Rev. } {\bfseries D 96} ((2017)  085006},
			[\href{https://doi.org/10.48550/arXiv.1706.03917}{\ttfamily arXiv:1706.03917}].	
			
			\bibitem{fan2103.04420}
			Wei Fan, Angelos Fotopoulos, Stephan Stieberger, Tomasz R. Taylor and Bin Zhu, \emph{Conformal Blocks from Celestial Gluon Amplitudes},
			\href{https://doi.org/10.1007/JHEP05(2021)170}{\emph{Journal of High Energy
					Physics} {\bfseries 2021} (2021) 170},
			[\href{https://doi.org/10.48550/arXiv.2103.04420}{\ttfamily
				arXiv:2103.04420}].
			
			\bibitem{Strominger2104.13432}
			Alexander Atanasov, Walker Melton, Ana-Maria Raclariu, Andrew Strominger, \emph{Conformal Block Expansion in Celestial CFT},
			{\emph{Phys. Rev.} {\bfseries D 104} (2021)   126033 },
			[\href{https://doi.org/10.48550/arXiv.2104.13432}{\ttfamily arXiv:2104.13432}].	
			
			\bibitem{Solodukhin0303006}		
			Jan de Boer and Sergey N. Solodukhin, \emph{A holographic reduction of Minkowski space-time}, {\emph{Nucl.Phys.} {\bfseries B665} (2003)
				 545-593}, [\href{https://doi.org/10.48550/arXiv.hep-th/0303006}{{\ttfamily
					arXiv:hep-th/0303006}}].	
			
			
				\bibitem{Solodukhin0405252}		
			Sergey N. Solodukhin, \emph{Reconstructing Minkowski Space-Time},  [\href{https://doi.org/10.48550/arXiv.hep-th/0405252}{{\ttfamily
					arXiv:hep-th/0405252}}].	
			
			
				\bibitem{Cheung}
			Clifford Cheung, Anton de la Fuente and Raman Sundrum, \emph{4D Scattering Amplitudes and Asymptotic Symmetries from 2D CFT}, 
			\href{https://doi.org/10.1007/JHEP01(2017)112}{\emph{Journal of
					High Energy Physics} {\bfseries 2017} (2017) 112},
			[\href{https://doi.org/10.48550/arXiv.1609.00732}{{\ttfamily arXiv:1609.00732}}].
			
				\bibitem{Strominger1905.09809}
			Adam Ball, Elizabeth Himwich, Sruthi A. Narayanan, Sabrina Pasterski, Andrew Strominger, \emph{Uplifting AdS3/CFT2 to Flat Space Holography}, 
			\href{https://doi.org/10.1007/JHEP08(2019)168}{\emph{Journal of
					High Energy Physics} {\bfseries 2019} (2019) 168},
			[\href{https://doi.org/10.48550/arXiv.1905.09809}{{\ttfamily arXiv:1905.09809}}].
			
			\bibitem{Strominger2204.10249}
			Eduardo Casali, Walker Melton, Andrew Strominger, \emph{Celestial Amplitudes as AdS-Witten Diagrams}, 
			\href{https://doi.org/10.1007/JHEP11(2022)140}{\emph{Journal of
					High Energy Physics} {\bfseries 2022} (2022) 140},
			[\href{https://doi.org/10.48550/arXiv.2204.10249}{{\ttfamily arXiv:.2204.10249}}].
			
			\bibitem{Strominger2312.07820}
			Walker Melton, Atul Sharma, Andrew Strominger,  \emph{Celestial Leaf Amplitudes},  [\href{https://doi.org/10.48550/arXiv.2312.07820}{{\ttfamily
					arXiv:2312.07820}}].	
			
				\bibitem{lizzi1986quantization}
			F.~Lizzi, B.~Rai, G.~Sparano and A.~Srivastava, \emph{Quantization of the null
				string and absence of critical dimensions}, {\emph{Physics Letters B}
				{\bfseries 182} (1986) 326--330}.
			
			\bibitem{gamboa1990null}
			J.~Gamboa, C.~Ramirez and M.~Ruiz-Altaba, \emph{Null spinning strings},
			{\emph{Nuclear Physics B} {\bfseries 338} (1990) 143--187}.
			
			
			\bibitem{lindstrom1991zero}
			U.~Lindstr{\"o}m, B.~Sundborg and G.~Theodoridis, \emph{The zero tension limit
				of the superstring}, {\emph{Physics Letters B} {\bfseries 253} (1991)
				319--323}.
				
			\bibitem{lindstrom19303173}
				U.~Lindstr{\"o}m,  \emph{The Zero Tension Limit of Strings and Superstrings},  [\href{https://doi.org/10.48550/arXiv.hep-th/9303173}{{\ttfamily
						arXiv:hep-th/9303173}}].	
			
			\bibitem{siegel2015amplitudes}
			W.~Siegel, \emph{Amplitudes for left-handed strings},
			\href{https://arxiv.org/abs/arXiv:1512.02569}{{\ttfamily arXiv:1512.02569}}.
			
			\bibitem{casali2016null}
			E.~Casali and P.~Tourkine, \emph{On the null origin of the ambitwistor string},
			\href{http://dx.doi.org/10.1007/JHEP11(2016)036}{\emph{Journal of High Energy
					Physics} {\bfseries 2016} (2016) 36},
			[\href{https://arxiv.org/abs/arXiv:1606.05636}{\ttfamily
					arXiv:1606.05636}].
				
				\bibitem{Strominger1910.07424} 	
			Monica Pate, Ana-Maria Raclariu, Andrew Strominger, Ellis Ye Yuan, \emph{Celestial Operator Products of Gluons and Gravitons},
			[\href{https://doi.org/10.48550/arXiv.1910.07424}{\ttfamily arXiv:1910.07424}].	
					
					\bibitem{Strominger2105.00331} 	
				Erin Crawley, Noah Miller, Sruthi A. Narayanan, Andrew Strominger, \emph{State-Operator Correspondence in Celestial Conformal Field Theory},
				\href{https://doi.org/10.1007/JHEP09(2021)132}{\emph{Journal of High Energy Physics} {\bfseries 2021} (2021)   132},
					[\href{https://doi.org/10.48550/arXiv.2105.00331}{\ttfamily arXiv:2105.00331}].	
				
				\bibitem{Sharma2111.02279} 	
				Tim Adamo, Wei Bu, Eduardo Casali, Atul Sharma, \emph{Celestial operator products from the worldsheet},
				\href{https://doi.org/10.1007/JHEP06(2022)052}{\emph{Journal of High Energy
						Physics} {\bfseries 2022} (2022) 052},
				[\href{https://doi.org/10.48550/arXiv.2111.02279}{\ttfamily
					arXiv:2111.02279}].
				
			
			\bibitem{bcfw} 	
			Ruth Britto, Freddy Cachazo, Bo Feng, Edward Witten, \emph{Direct Proof Of Tree-Level Recursion Relation In Yang-Mills Theory},
			 {\emph{Physical review letters} {\bfseries 94} (2005) 181602}, [\href{https://doi.org/10.48550/arXiv.hep-th/0501052}{{\ttfamily
					arXiv:hep-th/0501052}}].
				
				\bibitem{Taylor}
			Tomasz R. Taylor, \emph{A Course in Amplitudes}, {\emph{Physics Reports} {\bfseries  691} (2017)
				1-37},
			[\href{https://doi.org/10.48550/arXiv.1703.05670}{{\ttfamily
					arXiv:1703.05670}}].
				
				\bibitem{verlinde}
				Lauren McGough, Márk Mezei, Herman Verlinde, \emph{Moving the CFT into the bulk with $T\bar{T}$},	
					\href{https://doi.org/10.1007/JHEP04(2018)010}{\emph{Journal of High Energy
						Physics} {\bfseries 2018} (2018) 10},
				[\href{https://doi.org/10.48550/arXiv.1611.03470}{\ttfamily arXiv:1611.03470}].	
				
					\bibitem{Zamolodchikov0401146}
				Alexander B. Zamolodchikov, \emph{Expectation value of composite field $T\bar{T}$ in two-dimensional quantum field theory}, [\href{https://doi.org/10.48550/arXiv.hep-th/0401146}{\ttfamily arXiv:hep-th/0401146}].
					
					\bibitem{Zamolodchikov1608.05499}
					F.A. Smirnov, A.B. Zamolodchikov, \emph{On space of integrable quantum field theories}, 
					{\emph{Nuclear Physics } {\bfseries  B915} (2017) 363-383},
					[\href{https://doi.org/10.48550/arXiv.1608.05499}{\ttfamily arXiv:1608.05499}].
			
			\bibitem{shaw}
	John D. Barrow, Douglas J. Shaw, \emph{The Value of the Cosmological Constant}, {\emph{Gen Relativ Gravit} {\bfseries 43} (2011)
		2555–2560},
	[\href{https://doi.org/10.48550/arXiv.1105.3105}{{\ttfamily
			arXiv:1105.3105}}].
	
		\bibitem{Cardy1986}
	H. W. J. Bl$\ddot{\rm{o}}$te, John L. Cardy, and M. P. Nightingale, \emph{Conformal invariance, the central charge, and universal finite-size amplitudes at criticality},
	{\emph{Physical review letters} {\bfseries 56} (1986) 742},
	
	\bibitem{weinberg}
	Steven Weinberg, \emph{The cosmological constant problem}, {\emph{Rev. Mod. Phys.} {\bfseries  61-1} (1989)
		1-23}.
    
    	\bibitem{Witten9802150}
    Edward Witten,  \emph{Anti De Sitter Space And Holography}, {\emph{Adv.Theor.Math.Phys.} {\bfseries 2} (1998)
    	253-291}, [\href{https://doi.org/10.48550/arXiv.hep-th/9802150}{{\ttfamily
    		arXiv:hep-th/9802150}}].
    	  
   \bibitem{Mason1905.09224}
   Tim Adamo, Lionel Mason, Atul Sharma, \emph{Celestial amplitudes and conformal soft theorems}, {\emph{Class.Quant.Grav.} {\bfseries 36} (2019)
   	205018},
   [\href{https://arxiv.org/abs/1905.09224v3}{{\ttfamily
   		arXiv:1905.09224}}].
    
     \bibitem{Ruzziconi}	
    Adrien Fiorucci, Daniel Grumiller, Romain Ruzziconi, \emph{Logarithmic Celestial Conformal Field Theory}, {\emph{Physical Review D} {\bfseries 109} (2024) L021902},
    [\href{https://doi.org/10.48550/arXiv.2305.08913}{{\ttfamily
    		arXiv:2305.08913}}].
    	
    	 \bibitem{Song-He}
    	Song He, Pujian Mao, Xin-Cheng Mao, \emph{T$\bar{T}$ deformed soft theorem}, {\emph{Phys. Rev. D} {\bfseries 107} (2023) L101901},
    	[\href{https://doi.org/10.48550/arXiv.2209.01953}{\ttfamily
    			arXiv:2209.01953}].
    	
\end{thebibliography}
    \end{document}